\documentclass[aps,prb,twocolumn,floatfix,showpacs]{revtex4-1}
\usepackage{amsmath,amssymb,graphicx,bm}
\begin{document}
\title{Multi-orbital simplified parquet equations for strongly
  correlated electrons}
% \shorttitle{Multiorbital parquet equations}

\author{Pavel Augustinsk\'y} \author{V\'aclav Jani\v{s}}

\affiliation{Institute of Physics, Academy of Sciences of the Czech
  Republic, Na Slovance 2, CZ-18221 Praha 8, Czech Republic}
\email{janis@fzu.cz, august@fzu.cz}

\date{\today}

\begin{abstract}
  We extend an approximation earlier developed by us for the
  single-impurity Anderson model to a full-size impurity solver for
  models of interacting electrons with multiple orbitals. The
  approximation is based on parquet equations simplified by separating
  small and large energy fluctuations justified in the critical region
  of a pole in the two-particle vertex.  We show that an $l$-orbital
  model with most general interaction is described within this
  approximation by $4l^{2}\times 4l^{2}$ matrices and is Fermi liquid in
  the metallic phase.  We explicitly calculate properties of a
  paramagnetic solution of a two-orbital Hubbard model with a Hund
  exchange and orbital splitting within the dynamical mean-field
  approximation. We trace the genesis of a metal-insulator transition
  induced by a crystal field and vanishing of the Kondo
  quasiparticle peak in strongly correlated orbitally asymmetric systems.
\end{abstract}

\maketitle %newpage

\section{Introduction}\label{sec:Intro}

Understanding the effects of electron correlations is important not
only for comprehending the behavior of a wide class of materials. It
also allows one to get to fundamental principles of quantum
theory. Quantum dynamical effects caused by electron correlations are
still in the center of interest of condensed-matter
physicists. Although a lot of aspects of electron correlations have
been successfully explained, there are phenomena the picture of which
is incomplete or lacks substantial pieces. The main obstacle in
reaching a complete knowledge of the effects of electron correlations
is the difficulty of a reliable description of dynamical many-body
phenomena.

The major effort in grasping strong electron correlations is focused
on single-impurity Anderson model (SIAM).\cite{Anderson61} This model
was conceived to provide a simple picture of formation of local
magnetic moments. It appeared rather soon that formation of local
moments is connected with the Kondo effect.\cite{Kondo64} Since then
SIAM has been attracting standing attention of condensed-matter
theorists and experimentalists. The first wave of interest in SIAM was
crowned by finding an exact ground state of the model with a flat band
of conduction electrons for all interaction strengths.\cite{Tsvelik83}
The Kondo effect in SIAM was then explained and quantitatively
described. The exact expression for the Kondo asymptotics in SIAM has
become a hallmark and reference for dynamical effects of strong
electron correlations.
  
Impurity models suit for a description of quantum dots and
nano-particles\cite{Borda03,Choi05} but are unable to describe bulk,
macroscopic effects of electron correlations. It is, however,
generally believed that a local electron interaction in the
tight-binding scheme offers enough space for the description of most
of the dynamical effects of electron correlations. Success of the
dynamical mean-field theory (DMFT) of electron correlations has proved
this.\cite{Georges96} Dynamical mean-field approach, that is local
dynamical approximations, renewed interest in SIAM, since the basic
ingredient of DMFT is an impurity solver.  The latter essentially is
an impurity model with a self-consistent condition on the local
one-electron propagators. The exact thermodynamic solution of SIAM,
however, cannot be used, since it does not produce the necessary
dynamical renormalization of the energy spectrum. A new wave of
attempts to find approximate dynamical solutions of SIAM in the
strong-coupling regime arose.

Presently, the most comprehensive quantitative approaches producing
dynamical properties of the impurity models are quantum Monte Carlo
(QMC)\cite{Fye88} with its continuous-time
extension,\cite{Rubtsov05,Werner06a} exact diagonalization (ED) with a
discretized bath,\cite{Caffarel94} and the numerical renormalization
group (NRG).\cite{Wilson75,Krishnamurthy80} The first two approaches
aim at a precise thermodynamics of the impurity problem, while the
last one best reproduces low-temperature and low-energy scales of
impurity models. The first two approaches work naturally in the
Matsubara formalism and need a numerical analytic continuation to
determine spectral properties, while NRG can be straightforwardly
extended to dynamical and spectral quantities.\cite{Costi94, Bulla98}
The numerical renormalization group represents now-a-days the most
accurate quantitative low-temperature and low-energy dynamical
solution of impurity problems.\cite{Hewson93}

Apart from pure numerical methods a series of analytically controlled
approximations have been proposed. An approximation based on
low-frequency renormalizations of Fermi-liquid parameters is able to
reproduce the Kondo scale in SIAM and fits for heavy Fermion
systems.\cite{Hewson93a} It is not, however, suitable for the
description of critical instabilities of the Fermi liquid
state. Strong-coupling expansions based on the infinite-interaction
model \cite{Keiter70,Pruschke89} reproduce as well the Kondo scale but
fail to reproduce the Fermi liquid regime in the weak-coupling limit.

Standard truncated weak-coupling expansions fail to reproduce
correctly the strong-coupling regime. An improvement can be reached if
the expansion is applied on vertex functions.  Expansions for
two-particle vertices are mostly based on multiple scatterings of
electrons on electrons (T-matrix, TMA) or electrons on holes (random
phase, RPA). The non-renormalized multiple electron-hole scatterings
lead to an unphysical pole that must be removed by a self-consistent
treatment. Standardly introduced one-electron self-consistency
(fluctuation-exchange approximation) removes the pole but fails to
reproduce the Kondo scale.\cite{Bickers89} A static one-electron
self-consistency of the Hartree type was introduced in an effective
approximate scheme, called local-moment approach.\cite{Logan98} The
expansion for the polarization operator (RPA vertex) uses one-electron
propagators with a broken spin symmetry in this approximation. The
spin symmetry of the equilibrium thermodynamic state is then recovered
via a construction of the self-energy calculated from the polarization
operator. In this way the RPA pole is not crossed and the
strong-coupling limit keeps the polarization operator in the critical
region of the RPA pole. The spectral function displays the correct
Kondo asymptotics of the quasiparticle peak. This theory, however,
contains an artificial symmetry breaking and an auxiliary fitting
parameter, the length of the local magnetic moment, in the
strong-coupling regime.

An alternative way of introducing a self-consistency at the
two-particle level was proposed by us.\cite{Janis07a,Janis08} We
suggested to use parquet equations constructed from multiple
electron-electron and electron-hole scatterings and use the
criticality of the RPA pole for a simplification leading to a
manageable theory. This approximation allows us to control
analytically the RPA pole and the genesis of the Kondo
asymptotics. The derived Kondo behavior reproduces universal features
of the exact Bethe-ansatz solution.

Real materials are more complex and cannot be described by elementary
impurity models. One has to extend the methods to multi-orbital
Hamiltonians. Not all the methods used as impurity solvers for
single-orbital models work also in the multi-orbital case.  The exact
solution can be extended to $SU(n)$ models only in the limit
$U\to\infty$.\cite{Ogievetski83} There are natural extensions of
QMC\cite{Bonca93,Kotliar06,Werner06b}, ED,\cite{Liebsch05,Inaba05} and
NRG\cite{Galpin05,Bulla08} methods to multiple-orbitals. All the
numerically exact approaches are, however, computationally rather
expensive for realistic atomic structures. There is hence still a need
for simpler, computationally less expensive semi-analytic approaches
that could qualitatively correctly describe the transition from the
weak to the strong coupling regimes in multi-orbital impurity and
lattice models.

Some of the existing analytically controlled approaches to strong
electron correlations have recently been extended to multi-orbital
models, mostly $SU(n)$ Anderson models. Various spectral and
thermodynamic quantities have recently been calculated within the
renormalized Fermi-liquid expansion,\cite{Nishikawa10} fluctuation
exchange,\cite{Drchal05} or local-moment
approach.\cite{Galpin09,Kauch09} The aim of this paper is to present a
generalization of a two-particle approach based on simplified parquet
equations to multi-orbital models and create a full-scale impurity
solver for realistic models of correlated electrons. Unlike the most
of the other multi-orbital approaches, we formulate the method so that
it can directly be applied not only to impurity models but also to
lattice models within the dynamical mean-field theory. The
approximation is formulated in real frequencies and hence the spectral
properties are directly available. The low and high frequency behavior
of the one-electron spectral function, its three-peak structure, is
qualitatively well reproduced and the method allows one to control
analytically the critical behavior of Bethe-Salpeter equations for the
two-particle vertex.

The paper is organized as follows. We introduce a general
multi-orbital perturbation expansion for two-particle vertices in
Sec.~\ref{sec:Renormalized_PT}. In Sec.~\ref{sec:Simplified_MPE} we
derive simplified parquet equations with electron-hole and
electron-electron multiple scatterings and show how the self-energy is
calculated from the full two-particle vertex. We present explicit
equations to be solved in a two-orbital case in
Sec.~\ref{sec:Two_orbital}.  The derived approximation is used to
obtain a numerical solution of a two-orbital model with Hund's
coupling and orbital splitting in Sec.~\ref{sec:Results}.  We also
compare our results with those from other approaches there.
Reliability and domains of applicability of the derived approximation
are discussed in the final Section~\ref{sec:Conclusions}.

\section{Perturbation expansion for multi-orbital models of correlated
  electrons}\label{sec:Renormalized_PT}

\begin{figure}
  \includegraphics{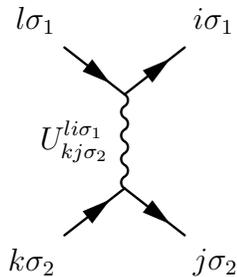}
  \caption{General electron-electron interaction in the notation we
    use in this paper. Index $\sigma$ stands for spin, while $i,j,k,l$
    comprise all other dynamical degrees of freedom characterizing the
    electron state. Conservation law in the dynamical variables $i + j
    = k + l$ holds.}\label{fig:Int-general}
\end{figure}

We start with a model Hamiltonian with multiple orbitals, where the
band structure is resolved and kinetic energy is diagonal in orbital
indices. The unperturbed Hamiltonian then reads
\begin{equation}\label{eq:H0}
  \widehat{H}_{0} = \sum_{\mathbf{k}}\sum_{\iota,\sigma}
  \epsilon_{\iota}(\mathbf{k})
  \tilde{c}^{\dagger}_{\iota\sigma}(\mathbf{k})
  \tilde{c}^{\phantom{\dagger}}_{\iota\sigma}(\mathbf{k})  
\end{equation}
where $\mathbf{k}$, $\sigma$, and $\iota$ are momentum, spin and
orbital index of valence electrons.  The interaction multi-orbital
Hamiltonian can generally be represented in the direct space as
\begin{multline}\label{eq:HI}
  \widehat{H}_{I} = \frac 14
  \sum_{\substack{\mathbf{x}_{1},\mathbf{x}_{2}\\
      \sigma_{1},\sigma_{2}}}\sum_{\substack{\iota,
      \upsilon,\kappa,\lambda\\ \iota\le \upsilon, \kappa\\ \kappa \le
      \lambda}}
  U^{\lambda\iota,\sigma_{1}}_{\kappa\upsilon,\sigma_{2}}(\mathbf{x}_{1}
  - \mathbf{x}_{2} ) c^{\dagger}_{\iota\sigma_{1}}(\mathbf{x}_{1})
  c^{\dagger}_{\upsilon\sigma_{2}}(\mathbf{x}_{2}) \\ \times
  c^{\phantom{\dagger}}_{\kappa\sigma_{2}}(\mathbf{x}_{2})
  c^{\phantom{\dagger}}_{\lambda\sigma_{1}}(\mathbf{x}_{1})
\end{multline}
where all possible inter-orbital transitions are
included. Orbital-momentum conservation poses a restriction on the
selection of orbital indices of the interaction matrix being $\iota +
\upsilon = \kappa + \lambda$.  We use a tight-binding approximation
taking the coordinate $\mathbf{x}$ from the lattice space. We will
assume the interaction to be local in the lattice space, that is,
Hubbard-like. A graphical representation of the interaction matrix is
plotted in Fig.~\ref{fig:Int-general}, where we used latin indices for
all the dynamical degrees of freedom characterizing the electron
states. They are generally four-momenta or Matsubara frequencies in
local approximations (dynamical mean-field theory) together with
orbital indices. We use this abbreviated notation to simplify
summations over intermediate states in the multi-orbital perturbation
expansion.

It is necessary to classify all possible interconnections of
interaction vertices via fermion lines for building up the complete
multi-orbital perturbation expansion. It is not an easy task and we
will sum only selected classes of two-particle diagrams. We use the
graphical representation of the interaction vertex from
Fig.~\ref{fig:Int-general} and choose the upper part of it so that
spin (not necessarily orbital index) is conserved along the horizontal
fermion lines. We choose electron propagation from left to right and
hole propagation from right to left.  We have three possibilities how
to connect two interaction vertices by two fermionic lines. They
correspond to three two-particle scattering channels. \cite{Janis99b}

\begin{figure}
  \includegraphics[width=8 cm]{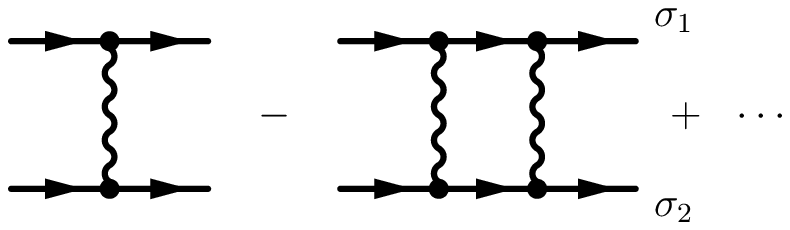}
  \caption{Connecting interactions via electron-electron
    propagation.}\label{fig:ee}
\end{figure}

The first possibility to connect interaction vertices is via
propagation of two electrons, that is, by parallel fermionic lines,
cf. Fig.~\ref{fig:ee}. Mathematically this process can be represented
via a matrix multiplication
%
% \begin{subequations}\label{eq:Channels}
\begin{equation}\label{eq:EE-Channel}
  \left( UGG\star U\right)^{li,\sigma_{1}}_{kj,\sigma_{2}} = \sum_{i'} U^{li',\sigma_{1}}_{kj',\sigma_{2}} G_{i'\sigma_{1}}G_{j'\sigma_{2}} U^{i'i,\sigma_{1}}_{j'j,\sigma_{2}}
\end{equation}
where the latin indices stand for all dynamical variables of the
electron (momentum, frequency and orbital index). They obey a
conservation law $j'= k + l - i'$.

\begin{figure}
  \includegraphics[width=8 cm]{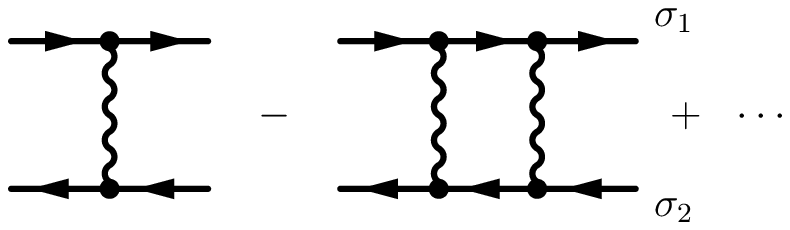}
  \caption{Connecting interactions via electron-hole
    propagation.}\label{fig:eh}
\end{figure}

Second connection of interaction vertices is via propagation of an
electron-hole pair, that is, by antiparallel fermionic lines. Here we
must distinguish two options: electron-hole scatterings and
polarization bubbles. The former process is plotted in
Fig.~\ref{fig:eh} and is represented via another matrix multiplication
\begin{equation}\label{eq:EH-Channel}
  \left( UGG\bullet U\right)^{li,\sigma_{1}}_{kj,\sigma_{2}} =
  \sum_{i'} U^{li',\sigma_{1}}_{k'k,\sigma_{2}}
  G_{i'\sigma_{1}}G_{j'\sigma_{2}} U^{i'i,\sigma_{1}}_{jk',\sigma_{2}} 
\end{equation}
where again the orbital-monentum conservation law $k'= i'+ k - l$
holds.

\begin{figure}
  \includegraphics[width=8 cm]{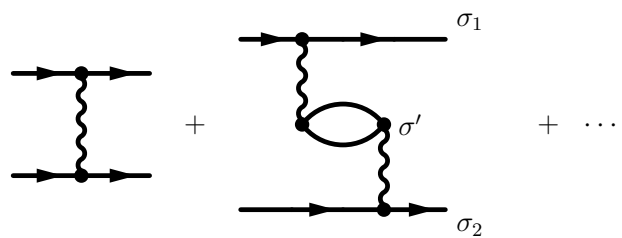}
  \caption{Connecting interactions via electron-hole triplet
    polarization bubbles (vertical channel).}\label{fig:v}
\end{figure}

A polarization bubble, shielding the interaction, is represented via a
matrix multiplication
\begin{equation}\label{eq:V-Channel}
  \left( UGG\circ U\right)^{li,\sigma_{1}}_{kj,\sigma_{2}} =
  \sum_{\sigma'}\sum_{j'} U^{li,\sigma_{1}}_{k'j',\sigma'}
  G_{k'\sigma'}G_{j'\sigma'} U^{k'j',\sigma'}_{kj,\sigma_{2}} 
\end{equation}
and is plotted in Fig.~\ref{fig:v}.  Notice that spin in this last
process is not conserved, but orbital-momentum conservation restricts
the intermediate indices of the dynamical variables to $k'= j'+ i -
l$.
% \end{subequations}


\begin{figure}
  \includegraphics[width=8 cm]{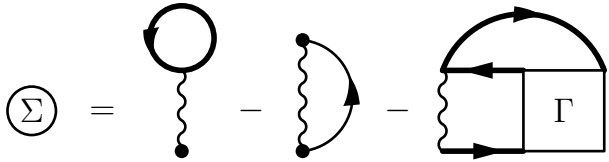}
  \caption{Schwinger-Dyson equation connecting the self-energy
    $\Sigma$ with vertex $\Gamma$. The first and the second term on
    the right-hand side of this equation are the Hartree and the Fock
    contribution.}\label{fig:SDE}
\end{figure}

We use these elementary scattering processes to construct the full
two-particle vertex $\Gamma$ from a two-particle perturbation
expansion. Once we have this vertex we construct the one-electron
self-energy as the fundamental quantity from which we derive all
thermodynamic properties. The self-energy is calculated from the
two-particle vertex via the Schwinger-Dyson equation, diagrammatically
represented in Fig.~\ref{fig:SDE}. Mathematically it reads
\begin{multline}\label{eq:SDE-general}
  \Sigma_{i\sigma} =
  \sum_{\sigma'}\sum_{j}U^{jj,\sigma'}_{ii,\sigma}G_{j\sigma'} -
  \sum_{j}U^{ji,\sigma}_{ij,\sigma} G_{j\sigma} \\
  - \sum_{\sigma'}\sum_{j',k'}
  U^{k'l',\sigma'}_{ij',\sigma}G_{j'\sigma} G_{k'\sigma'}
  G_{l',\sigma'} \Gamma^{l'k',\sigma'}_{j'i,\sigma} \ .
\end{multline}
Again the conservation law for intermediate variables $l'= i + k'- j'$
holds. The one-electron propagators in the perturbation expansion can
either be chosen as fully dynamically renormalized as suggested by
Baym and Kadanoff,\cite{Baym62} or only statically renormalized via
the Hartree approximation securing that the particle densities do not
change during the summation of diagrams. We discussed earlier that the
latter option is more appropriate in the strong-coupling regime, since
we can better control the two-particle singularity in the
electron-hole correlation function and consequently both low and high
energy scales are qualitatively better reproduced than in approximate
schemes with the fully renormalized propagators.\cite{Janis07a}

\section{Simplified parquet equations: matrix
  formulation} \label{sec:Simplified_MPE}

The generic function for our perturbation expansion is the
two-particle vertex $\Gamma$. Its approximate expression together with
the Schwinger-Dyson equation \eqref{eq:SDE-general} define our
approximate scheme. We hence start with simple approximations on
vertex $\Gamma$. We resort our analysis only to local approximations,
that is, we stay within the dynamical mean-field theory with only
local fluctuations and Matsubara frequencies, orbital index and spin
as dynamical variables. We will treat these degrees of freedom
separately.

The simplest approximations for vertex $\Gamma$, beyond the bare
interaction, are multiple scatterings of the same form as defined by
elementary processes from the preceding section. Sums of all multiple
scatterings of the same type are solutions to Bethe-Salpeter
equations. They can be represented with the appropriate matrix
multiplications. For electron-hole and electron-electron multiple
scatterings we obtain
\begin{align}\label{eq:BS_bare-eh}
  \widehat{\Phi}(i\nu_{m}) = \widehat{U} - \widehat{U \chi}(i\nu_{m})
  \bullet \widehat{\Phi}(i\nu_{m})\ , \\ \label{eq:BS_bare-ee}
  \widehat{\Psi}(i\nu_{m}) = \widehat{U} - \widehat{U \psi}(i\nu_{m})
  \star \widehat{\Psi}(i\nu_{m}) \ ,
\end{align}
respectively. The kernels of these equations are composed from the
bare interaction and either the electron-hole bubble or the
two-electron propagator. The kernel in the electron-hole channel
explicitly reads
\begin{equation}\label{eq:Uchi}
  \left(U\chi\right)^{\lambda\iota,\sigma_{1}}_{\kappa\upsilon\sigma_{2}}
  = U^{\lambda\iota,\sigma_{1}}_{\kappa\upsilon,\sigma_{2}}
  \chi^{\iota\sigma_{1}}_{\kappa\sigma_{2}}(i\nu_{m}) \\  
\end{equation}
with the electron-hole bubble
\begin{align} \label{eq:chi}
  \chi^{\iota\sigma_{1}}_{\kappa\sigma_{2}}(i\nu_{m}) &= \frac 1\beta
  \sum_{n}G_{\iota\sigma_{1}}(i\nu_{m} + i\omega_{n})
  G_{\kappa\sigma_{2}}(i\omega_{n}) \ .
\end{align}
Analogously we construct the kernel of the Bethe-Salpeter equation
\eqref{eq:BS_bare-ee} with multiple electron-electron scatterings. The
two-electron propagator then is
\begin{align}\label{eq:psi}
  \psi^{\iota\sigma_{1}}_{\upsilon\sigma_{2}}(i\nu^{\prime}_{m}) &=
  \frac 1\beta \sum_{n}G_{\iota\sigma_{1}}( i\nu^{\prime}_{m} -
  i\omega_{n}) G_{\upsilon\sigma_{2}}(i\omega_{n})\ .
\end{align}
We denote fermionic and bosonic Matsubara frequencies $\omega_{n}= (2n
+ 1)\pi T$ and $\nu_{m}=2m\pi T$, respectively (Boltzmann constant
$k_{B} =1$). Notice that the transfer frequency $\nu_{m}$, conserved
in electron-hole scatterings, is different from that conserved in the
electron-electron scattering channel, $\nu^{\prime}_{m}$. If the
incoming and outgoing frequencies of one electron are $\omega_{n}$ and
$\omega_{n'}$, respectively, we have $\nu^{\prime}_{m} = \nu_{m}+
\omega_{n} - \omega_{n'}$.

We do not take into account the third channel with triplet
polarization bubbles. There are two reasons why we can afford to do
this. First, the effects of the scatterings in the vertical channel
are similar to those from the electron-hole channel, at least
qualitatively. We hence do not lose track of any relevant effect in
the strong-coupling regime. Second, taking into account only two
scattering channels simplifies significantly approximate schemes for
the two-particle vertex.\cite{Janis09}

From the solutions of the Bethe-Salpeter equations for
$\widehat{\Phi}$ and $\widehat{\Psi}$ we obtain the full two-particle
vertex
\begin{multline}\label{eq:parguet-bare}
  \widehat{\Gamma}(i\nu_{m},i\nu_{m}+ i\omega_{n}- i\omega_{n'})\\ =
  \widehat{\Phi}(i\nu_{m}) + \widehat{\Psi}(i\nu_{m}+ i\omega_{n}-
  i\omega_{n'}) - \widehat{U}\ .
\end{multline}
Such a vertex defines the self-energy of the so-called FLEX
approximation if the one-electron propagators in the Bethe-Salpeter
equations \eqref{eq:BS_bare-eh} and \eqref{eq:BS_bare-ee} are fully
renormalized, that is, they contain the self-energy determined by the
approximate vertex $\Gamma$ from Eq.~\eqref{eq:parguet-bare}. Such an
approximation is known to be unreliable in the strong-coupling
regime.\cite{Janis99b}

Intermediate electron coupling is marked by proximity of a pole in the
electron-hole vertex $\Phi$.  If the transfer frequency $\nu=0$ then
we can define a small dimensionless scale
\begin{equation}
  a = 1+{\rm min}\ \text{Sp}\left[\widehat{U^{\sigma_{1}}_{\sigma_2} \chi^{\sigma_{1}}_{\sigma_2}}\right]
\end{equation}
approaching zero in the random-phase approximation with $\Gamma =
\Phi$, a solution of Eq.~\eqref{eq:BS_bare-eh}, when $U\to U_{c}$. We
used $\text{Sp}\left[ \widehat{A}\right]$ to denote the set of
eigenvalues of matrix $\widehat{A}$. This pole is unphysical and
indicates that the perturbation expansion for $U\ge U_{c}$ must be
renormalized. The critical interaction $U_{c}$ from RPA defines an
upper limit on the weak-coupling regime where perturbation theory with
selected classes of generic diagrams can safely be applied. We speak
about strong coupling if $U > U_{c}$, where only self-consistent
approximations may be reliable.

One, and most often used option how to suppress the unphysical pole,
or better to shift it to infinite interaction strength is to employ
one-particle dynamical self-consistency.  We use the self-energy
calculated from the Schwinger-Dyson equation \eqref{eq:SDE-general}
with vertex $\Gamma$ from Eq.~\eqref{eq:parguet-bare} to renormalize
the one-electron propagators in Bethe-Salpeter equations
\eqref{eq:BS_bare-eh} and \eqref{eq:BS_bare-ee}. This one-particle
self-consistency succeeds in shifting the unphysical pole to infinity,
but the Kondo regime defined as $a\to 0$ is not properly
reproduced. Moreover, FLEX-type approximations completely smear out
the large-frequency structures. There are no satellite Hubbard
bands.\cite{Janis99b}

A better systematic alternative how to move the RPA critical
interaction $U_{c}$ to infinity is to introduce a two-particle
self-consistency. The most straightforward way to do so is to use the
parquet approach. Its basic idea is to replace the bare interaction in
Bethe-Salpeter equations \eqref{eq:BS_bare-eh} and
\eqref{eq:BS_bare-ee} by irreducible vertex functions $\Lambda^{eh}$
and $\Lambda^{ee}$, respectively. In the exact theory then
$\Phi\left[\Lambda^{eh}\right] = \Psi\left[\Lambda^{ee}\right] =
\Gamma$ and equation~\eqref{eq:parguet-bare} transforms to $\Gamma + U
= \Lambda^{eh} + \Lambda^{ee}$. When we combine the two Bethe-Salpeter
equations with the parquet equation~\eqref{eq:parguet-bare} modified
as above we obtain a set of non-linear integral equations for the
irreducible vertices $\Lambda^{eh}\left[U, G\right],
\Lambda^{ee}\left[U, G\right]$ and finally also for the full vertex
$\Gamma\left[U, G\right]$. It is, however, impossible to solve the
parquet equations in the strong-coupling regime with a vanishing Kondo
scale $a\to 0$. The problem lies in that all the vertex functions,
solutions of the full parquet equations, generally depend on three
frequencies and there is presently no method, either analytic or
numerical, that would predict the full analytic structure of the
two-particle vertices.  We hence have to approximate the full parquet
equations.

We proposed a simplification of the parquet equations in the Kondo
regime of the single-impurity Anderson model.\cite{Janis07a,Janis08}
This simplification utilizes a partial separation of large and small
frequency fluctuations in the Kondo regime when the solution of the
Bethe-Salpeter equation in the electron-hole channel is almost
singular.  In the non-renormalized perturbation expansion it is the
RPA pole. The principal idea of our simplification is that we separate
large fluctuations diverging at the critical point from those
remaining finite. In this way we describe correctly universal
quantities connected with the critical point in the electron-hole
Bethe-Salpeter equation. We neglect all finite fluctuations and keep
only the relevant ones diverging at the critical point. We then
replace all finite frequency-dependent functions by constants and keep
only the frequency dependence in the variables controlling large
fluctuations. This simplification is kind of a mean-field (simplest)
approximation for the dynamics in the strong-coupling regime.

We know from the analysis of multiple electron-electron and
electron-hole scatterings that only the latter can cause a singularity
in the vertex function. It means that the irreducible vertex in the
electron-hole channel $\Lambda^{eh}$ remains finite. We hence can
replace it by an effective interaction that we denote
$\overline{U}$. The renormalized Bethe-Salpeter equation in the
electron-hole channel is RPA-like with the following replacement
\begin{equation}\label{eq:Effective-interaction}
  \widehat{U} \longrightarrow \widehat{\overline{U}}, \qquad
  \widehat{\Phi}(i\nu_{m}) \longrightarrow\
  \widehat{\overline{\Phi}}(i\nu_{m})\ . 
\end{equation}
The effective interaction $\overline{U}$ will be determined from the
Bethe-Salpeter equation in the electron-electron channel by using the
parquet self-consistency. Within the scope of our approximation the
Bethe-Salpeter equation in the electron-electron scattering channel
holds only in an averaged form smearing out finite frequency
differences.  We proposed in Ref.~\onlinecite{Janis08} a decoupling of
the Bethe-Salpeter equation in the electron-electron scattering
channel so that an integral equation is converted to an algebraic,
solvable one. In this approximate simplification an averaged integral
kernel is assumed to diagonalize the integral equation. Such a
decoupling is not uniquely defined and should be chosen so that the
effective interaction is a real self-adjoint (positive definite)
matrix.  An equation for the effective interaction derived from a
decoupling with the right-averaged integral kernel used in the
single-orbital case reads
\begin{equation}\label{eq:EI-right}
  \widehat{\overline{U}\psi} = \widehat{U\psi} -
  \widehat{L}\star\left[\widehat{1} +
    \widehat{L}\star\right]^{-1}\widehat{L}  
\end{equation}
where we denoted
\begin{multline}\label{eq:L-function}
  L^{\lambda\iota,\sigma_{1}}_{\kappa\upsilon,\sigma_{2}}(i\omega_{n})
  \\ = \frac 1{\beta} \sum_{n'}
  \Lambda^{\lambda\iota,\sigma_{1}}_{\kappa\upsilon,\sigma_2}(i\nu_{-n-n'})
  G_{\iota\sigma_1}(i\omega_{n'}) G_{\upsilon\sigma_2}(i\omega_{-n'})\
  .
\end{multline}
This one-sided averaging in the multi-orbital case can lead to a
matrix of the effective interaction that is not self-adjoint. To avoid
this spurious behavior we introduce also a left-averaged integral
kernel in the electron-electron Bethe-Salpeter equation. We then
define another equation for the effective interaction
\begin{equation}\label{eq:EI-left}
  \widehat{\psi \overline{U}} = 
  \ \widehat{\psi U}   - \widehat{\mathcal{L}}\star\left[\widehat{1} +
    \widehat{\mathcal{L}}\star\right]^{-1}\widehat{\mathcal{L}} \ , 
\end{equation}
with a left-averaged integral kernel
\begin{multline}\label{eq:calL-function}
  {\mathcal{L}}^{\lambda\iota,\sigma_{1}}_{\kappa\upsilon,\sigma_{2}}(i\omega_{n})
  \\ = \frac 1{\beta} \sum_{n'} G_{\lambda\sigma_1}(i\omega_{n'})
  G_{\kappa\sigma_2}(i\omega_{-n'})
  \Lambda^{\lambda\iota,\sigma_{1}}_{\kappa\upsilon,\sigma_2}(i\nu_{-n-n'})
  \ .
\end{multline}
We demand that the matrix of the effective interaction
$\widehat{\overline{U}}$ be self-adjoint. It is always the case if we
use a sum of equations \eqref{eq:EI-right} and \eqref{eq:EI-left} to
determine the effective interaction.  That is, we choose an equation
symmetric with respect to right and left averaging of the integral
kernel of the Bethe-Salpeter equation in the nonsingular
electron-electron channel.

The dominant contribution to the integral kernel in the
electron-electron scattering channel comes from the singular part of
the vertex from the electron-hole scattering channel, being
$\widehat{\Lambda} = \widehat{\Phi} - \widehat{U}$. Since the averaged
integral kernels $L$ and $\mathcal{L}$ depend on a fermionic Matsubara
frequency, we assume that Eqs.~\eqref{eq:EI-right}, \eqref{eq:EI-left}
are valid only for the lowest frequency, being zero at zero
temperature. At finite temperatures we use the lowest or a few low
frequencies, symmetric around zero, and add all equations for these
frequencies. The effective interaction is then determined form an
average over Eq.~\eqref{eq:EI-right} and Eq.~\eqref{eq:EI-left} for a
few lowest-lying fermionic frequencies. The low-temperature
asymptotics is in this way well reproduced. The higher the temperature
the more Matsubara frequencies we must take into account to maintain
accuracy of the zero-temperature solution. We know, however, that
thermal fluctuation suppress the Kondo resonance and smear out the RPA
singularity in the electron-hole correlation function.\cite{Janis08}

Equations \eqref{eq:BS_bare-eh},
\eqref{eq:EI-right}-\eqref{eq:calL-function} are our simplified
parquet equations determining the two-particle vertices
$\overline{\Phi}$ and $\overline{U}$. The full two-particle vertex
$\Gamma$ is determined from equation~\eqref{eq:parguet-bare}. Within
our approximation it reduces to the renormalized RPA vertex
$\overline{\Phi}$.  We use it to determine the self-energy. The
respective Schwinger-Dyson equation reads
\begin{widetext}
  \begin{multline}\label{eq:SDE-SPE}
    \Sigma_{\iota\sigma}(i\omega_{n}) = \frac 1\beta
    \sum_{m}\sum_{\sigma'\upsilon}U^{\upsilon\upsilon,\sigma'}_{\iota\iota,\sigma}G_{\upsilon\sigma'}(i\omega_{n
      + m}) - \frac 1\beta
    \sum_{m}\sum_{\upsilon}U^{\upsilon\iota,\sigma}_{\iota\upsilon,\sigma}
    G_{\upsilon\sigma}(i\omega_{n + m})  \\
    - \frac 1{\beta^{2}}
    \sum_{n',m}\sum_{\sigma'}\sum_{\iota',\kappa'}
    U^{\kappa'\lambda',\sigma'}_{\iota\iota',\sigma}G_{\iota'\sigma}(i\omega_{n+m})
    G_{\kappa'\sigma'}(i\omega_{n'})
    G_{\lambda',\sigma'}(i\omega_{n'-m})
    \overline{\Phi}^{\lambda'\kappa',\sigma'}_{\iota'\iota,\sigma}(i\nu_{m})\
    ,
  \end{multline}
\end{widetext}
where conservation of the orbital index $\lambda'= \iota + \kappa'-
\upsilon'$ holds.

The approximation is almost complete. The last thing we have to do is
to specify the one-electron propagators in the parquet equations for
the two-particle vertices. We know that the full one-particle
self-consistency where the one-electron propagators are renormalized
by the self-energy from Eq.~\eqref{eq:SDE-SPE} does not lead to the
Kondo scale as $U\to\infty$ and also smears out the satellite Hubbard
bands.\cite{Janis08} We hence use the Hartree one-electron propagators
that in the dynamical mean-field theory is
\begin{multline}\label{eq:1GF-general}
  G_{\iota\sigma}(i\omega_{n}) \\ = \int\frac {d\epsilon
    \rho_{\iota}(\epsilon)} {i\omega_{n} + \mu_{\iota} + \sigma h -
    \epsilon - \frac 12 \sum_{\lambda,\sigma'}
    U^{\lambda\lambda,\sigma'}_{\iota\iota,\sigma}n_{\lambda\sigma'}}
\end{multline}
and for the impurity Anderson model
\begin{multline}\label{eq:1GF-SIAM}
  G_{\iota\sigma}(i\omega_{n}) \\ = \frac 1{i\omega_{n} + \mu_{\iota}
    + \sigma h - \Delta(i\omega_{n}) - \frac 12 \sum_{\lambda,\sigma'}
    U^{\lambda\lambda,\sigma'}_{\iota\iota,\sigma}n_{\lambda\sigma'}}
  \ .
\end{multline}
We denoted $\mu_{\iota} = \mu - E_{\iota}$, where $E_{\iota}$ is the
center of the $\iota$'s orbital and $\mu$ is the chemical potential
fixing the total charge density. The Hartree propagators are
statically self-consistent, that is, the particle densities
$n_{\lambda \sigma}$ are determined from the fully renormalized
propagator
\begin{equation}\label{eq:1P-density}
  n_{\lambda\sigma} = \frac 1\beta
  \sum_{n}G_{\lambda\sigma}\left(i\omega_{n} -
    \Delta\Sigma_{\lambda\sigma}(i\omega_{n})\right)\ . 
\end{equation}
We denoted $\Delta\Sigma$ the dynamical part of the self-energy, a
correction to the Hartree term. The chemical potential is then
adjusted so that the total particle density is fixed. The approximate
scheme based on the simplified parquet equations is now complete. We
demonstrate explicitly applicability of this approximation on a
two-orbital model in the next sections.

\section{Two-orbital model}\label{sec:Two_orbital}

\begin{widetext}

  \begin{figure}
    \includegraphics{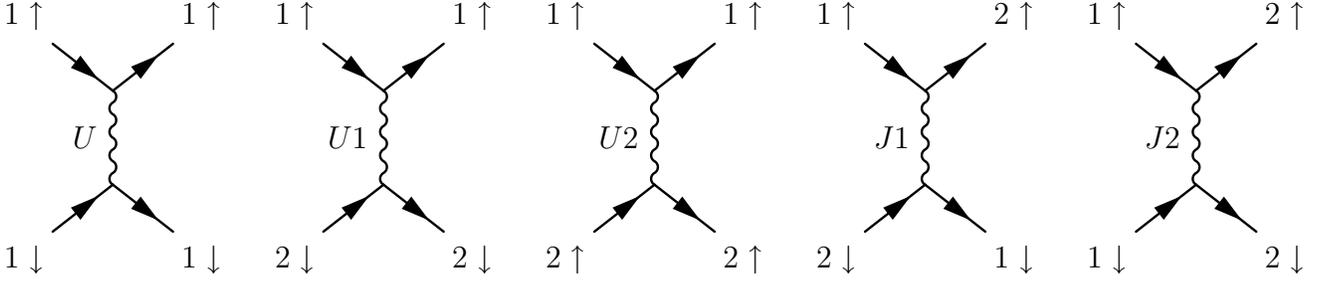}
    \caption{Different elementary scattering processes in the
      two-orbital model. Each of these processes is renormalized
      independently.}\label{fig:Mult_int}
  \end{figure}

\end{widetext}

We use a general two-orbital interacting Hamiltonian comprising also
Hund's exchange coupling\cite{Imada98}
\begin{multline}\label{eq:Hamiltonian-2O}
  H_I = U \sum_\iota n_{\iota\uparrow} n_{\iota\downarrow} +
  \sum_{\sigma\sigma'} \left[U_{1}\eta_{\sigma\sigma'} +
    U_{2}\delta_{\sigma\sigma'}\right] n_{1\sigma} n_{2\sigma'}\\
  -
  \left(J_{1}c^{\dagger}_{2\uparrow}c^{\dagger}_{1\downarrow}c^{\phantom{\dagger}}_{2\downarrow}c^{\phantom{\dagger}}_{1\uparrow}
    +
    J_{2}c^{\dagger}_{2\uparrow}c^{\dagger}_{2\downarrow}c^{\phantom{\dagger}}_{1\downarrow}
    c^{\phantom{\dagger}}_{1\uparrow}+H.c.\right)
\end{multline}
where $\eta_{\sigma\sigma'} + \delta_{\sigma\sigma'} =1$. A graphical
representation of single terms contributing to this two-orbital
Hamiltonian is presented in Fig.~\ref{fig:Mult_int}. Each input
parameter of this Hamiltonian will be renormalized independently.

Only one of the bare interacting terms denoted $U_{2}$ is spin
triplet.  This term does not mix with the other scattering processes
in the Hamiltonian from Eq.~\eqref{eq:Hamiltonian-2O}. The simplified
parquet equations for this term are identical with those from the
single-orbital case.  The effective interaction
$\overline{U}^{\sigma}_{2}$ is
\begin{equation}\label{eq:U2-bar}
  \overline{U}^{\sigma}_{2} = U_{2} -
  \frac{\left(L^{11,\sigma}_{22,\sigma}\right)^2}{\left(1+L^{11,\sigma}_{22,\sigma}\right)\psi^{1\sigma}_{2\sigma}} 
\end{equation}
with the averaged integral kernel from the Bethe-Salpeter equation in
the electron-electron channel
\begin{multline}\label{eq:L_upup}
  L^{11,\sigma}_{22,\sigma}(i\omega_n) \\ = \frac 1{\beta}
  \sum_{n'}G_{1\sigma}(i\omega_{n'}) G_{2\sigma}(i\omega_{-n'})
  \Lambda^{11,\sigma}_{22,\sigma}(i\nu_{-n -n'})\ .
\end{multline}
The dynamical part of the (reducible) vertex from the electron-hole
scattering channel is
\begin{equation}\label{eq:Lambda_upup}
  \Lambda^{11,\sigma}_{22,\sigma}(i\nu_{m}) = -
  \frac{\left(\overline{U}^{\sigma}_{2}\right)^{2}
    \chi^{1\sigma}_{2\sigma}(i\nu_{m})}{1+\overline{U}_{2}
    \chi^{1\sigma}_{2\sigma}(i\nu_{m})}\ . 
\end{equation}

The other terms in the interaction Hamiltonian from
Eq.~\eqref{eq:Hamiltonian-2O} mix in the parquet equations and must be
treated simultaneously. Since only specific inter-orbital processes
are allowed we can simplify the representation of the interaction term
from the preceding section and reduce super-matrices (tensors) with
four indices to regular matrices with two indices.  A relation between
matrix indices and super-indices can be chosen as follows
$1\equiv(1,1)$, $2\equiv(2,2)$, $3\equiv(1,2)$, $4\equiv(2,1)$. If we
go over from super-matrices to matrices we have to distinguish two
representations corresponding to two matrix multiplications used in
constructing the Bethe-Salpeter equations. The interaction matrix
suitable for summation of multiple electron-electron scatterings is
\begin{equation}
  % \[
  \widehat{U}^{ee} =\left(
    \begin{array}{cccc}
      U & J_2 & 0& 0 \\
      J_2 & U & 0 &0 \\
      0 & 0 & U_1 & J_1\\
      0 & 0 & J_1 & U_1
    \end{array}
  \right)\ .
  % \]
\end{equation}
The interaction matrix for the electron-hole multiple scatterings
interchanges the two Hund couplings $J_{1}$ and $J_{2}$. Hence we have
\begin{equation}
  % \[
  \widehat{U}^{eh} = \left(
    \begin{array}{cccc}
      U & J_1 & 0 & 0 \\
      J_1 & U & 0 & 0 \\
      0 & 0 & U_1 & J_2\\
      0 & 0 & J_2 & U_1
    \end{array}
  \right)\ .
  % \]
\end{equation}
Each matrix representation is chosen so that the scattering processes
in the respective channel are simple matrix multiplications.

The matrix integral kernel for the Bethe-Salpeter equation in the
electron-hole channel is
\begin{equation}
  \widehat{K} \equiv \widehat{\overline{U} \chi} =
  \left(
    \begin{array}{cccc}
      \overline{U}_{01} \chi^{1}_{1} & \overline{J}_1 \chi^{2}_{2} & 0 & 0 \\
      \overline{J}_1 \chi^{1}_{1} & \overline{U}_{02} \chi^{2}_{2} & 0  & 0 \\
      0 & 0 & \overline{U}_1\chi^{1}_{2}  & \overline{J}_2 \chi^{2}_{1}\\
      0 & 0 & \overline{J}_2 \chi^{1}_{2} & \overline{U}_1 \chi^{2}_{1}
    \end{array}
  \right),
\end{equation}
while the right-averaged kernel in the Bethe-Salpeter equation in the
electron-electron channel reads
\begin{equation}
  \widehat{L}^{ee} =
  \left(
    \begin{array}{cccc}
      L^{11}_{11} &L^{12}_{12}  & 0 & 0 \\[1pt]
      L^{21}_{21} & L^{22}_{22} & 0 & 0\\
      0 & 0 & L^{11}_{22} & L^{12}_{21}\\[1pt]
      0 & 0 &  L^{21}_{12} & L^{22}_{11}
    \end{array}
  \right)\ ,
\end{equation}
with
\begin{multline}
  L^{\lambda\iota,\sigma_{1}}_{\kappa\upsilon,\sigma_{2}}(i\omega_n)\\
  = \frac 1{\beta} \sum_{n'}
  \Lambda^{\lambda\iota,\sigma_{1}}_{\kappa\upsilon,\sigma_{2}}(i\nu_{-n
    -n'}) G_{\iota\sigma_{1}}(i\omega_{n'})
  G_{\upsilon\sigma_{2}}(i\omega_{-n'})
\end{multline}
and $\widehat{\Lambda} = \widehat{\overline{\Phi}} -
\widehat{\overline{U}}$. Analogously we construct the left-averaged
kernel $\mathcal{L}$, if necessary. The matrix form of the simplified
parquet equations then is in the electron-hole channel
\begin{align}\label{eq:SPE_2O-EH}
  \widehat{\overline{\Phi}}& = \left[\widehat{1}+
    \widehat{K}\bullet\right]^{-1} \widehat{\overline{U}}^{eh}\
\end{align}
and in the electron-electron channel with right and left-averaged
integral kernels
\begin{multline}\label{eq:SPE_2O-EE}
  \widehat{\overline{U}\psi} + \widehat{\psi\overline{U}} =
  \widehat{U\psi}
  - \widehat{L}^{ee}\star\left[\widehat{1} + \widehat{L}^{ee}\star\right]^{-1}\widehat{L}^{ee} \\
  +\ \widehat{\psi U} -
  \widehat{\mathcal{L}}^{ee}\star\left[\widehat{1} +
    \widehat{\mathcal{L}}^{ee}\star\right]^{-1}\widehat{\mathcal{L}}^{ee}
  \ .
\end{multline}

All matrices appearing in the parquet equations~\eqref{eq:SPE_2O-EH}
and \eqref{eq:SPE_2O-EE} are block diagonal. We denote
\begin{align}
  \overline{\Phi} = \left(
    \begin{array}{cc}
      \overline{\Phi}_{\rm upp}  &  0\\
      0 &  \overline{\Phi}_{\rm low}
    \end{array}
  \right) \ , & \quad \overline{U} = \left(
    \begin{array}{cc}
      \overline{U}_{\rm upp}  & 0 \\
      0   &  \overline{U}_{\rm low}
    \end{array}
  \right)
\end{align}
and solve the parquet equations for each block separately. The
solution in the electron-hole scattering channel has the following
representation
\begin{widetext}
  \begin{align}
    \overline{\Phi}_{\rm upp} &= \frac{1}{1^{\phantom{\big(}}+
      \overline{U}_{01}\chi^{1}_{1} + \overline{U}_{02}\chi^{2}_{2} +
      \left(\overline{U}_{01}\overline{U}_{02} -
        \overline{J}_{1}^{2}\right)\chi^{1}_{1} \chi^{2}_{2}
    } %\times \nonumber\\ &\times
    \left(
      \begin{array}{cc}
        \overline{U}_{01}  + \left( \overline{U}_{01}\overline{U}_{02}
          - \overline{J}_{1}^{2}\right)\chi^{2}_{2} \quad&
        \overline{J}_{1} \\ 
        \overline{J}_{1}\quad & \overline{U}_{02} + \left(
          \overline{U}_{01}\overline{U}_{02} -
          \overline{J}_{1}^{2}\right) \chi^{1}_{1} 
      \end{array}
    \right)\
    \label{Phi_EH_uppeL^{ee}_block_Ubar}
  \end{align}
  and
  \begin{align}
    \overline{\Phi}_{\rm low} &= \frac{1}{1^{\phantom{\big(}}+
      \overline{U}_{1}\left(\chi^{1}_{2} + \chi^{2}_{1}\right) +
      \left(\overline{U}_{1}^{2} -
        \overline{J}_{2}^{2}\right)\chi^{1}_{2} \chi^{2}_{1}
    } %\times \nonumber\\ &\times
    \left(
      \begin{array}{cc}
        \overline{U}_{1}  + \left( \overline{U}_{1}^{2} - \overline{J}_{2}^{2}\right)\chi^{2}_{1} \quad& \overline{J}_{2} \\
        \overline{J}_{2}\quad & \overline{U}_{1} + \left( \overline{U}_{1}^{2} - \overline{J}_{2}^{2}\right) \chi^{1}_{2}
      \end{array}
    \right)\ .
    \label{Phi_EH_loweL^{ee}_block_Ubar}
  \end{align}
  Notice that the Hubbard interaction $U$ splits into two renormalized
  values $\overline{U}_{01}$ and $\overline{U}_{02}$ if the bubbles in
  different orbitals are different, $\chi^{1}_{1} \neq
  \chi^{2}_{2}$. Note that we suppressed the frequency index at the
  electron-hole bubble leading to a frequency dependence of vertex
  $\widehat{\overline{\Phi}}$.

  We now choose the singular part of the renormalized electron-hole
  vertex $\widehat{\Lambda} = \widehat{\overline{\Phi}} -
  \widehat{\overline{U}}$ and use it to obtain a solution of the
  Bethe-Salpeter equation in the electron-electron channel with an
  averaged (right, left) kernel that determines the matrix of the
  effective (renormalized) interaction. Its upper matrix block reads
  \begin{multline}\label{eq:EI-O2-upp}
    \widehat{\overline{U}_{upp}\psi} \equiv \left(
      \begin{array}{cc}
        \overline{U}_{01} \psi^{1}_{1} & \overline{J}_{2}\psi^{2} _{2}\\
        \overline{J}_{2}\psi^{1}_{1} & \overline{U}_{02} \psi^{2}_{2}
      \end{array}
    \right) = \left(
      \begin{array}{cc}
        U  \psi^{1}_{1} & J_2  \psi^{2}_{2}\\[1pt]
        J_2  \psi^{1}_{1} & U  \psi^{2}_{2}
      \end{array}
    \right)  - \frac{1}{(1 + L^{ee}_{11})(1 + L^{ee}_{22}) - L^{ee}_{12} L^{ee}_{21}} \\
    \times \left(
      \begin{array}{cc}
        \left(1 + L^{ee}_{22}\right)\left( L^{ee}_{11}\right)^{2} +
        L^{ee}_{12} L^{ee}_{21} \left( 1 - L^{ee}_{11}\right) \quad &  
        L^{ee}_{12}\left( L^{ee}_{11} + L^{ee}_{22} +  L^{ee}_{11} L^{ee}_{22} - L^{ee}_{12} L^{ee}_{21}\right)  
        \\[1pt]     
        L^{ee}_{21}\left( L^{ee}_{11} + L^{ee}_{22} + L^{ee}_{11} L^{ee}_{22} - L^{ee}_{12} L^{ee}_{21}\right)    \quad & 
        \left(1 + L^{ee}_{11}\right) \left(L^{ee}_{22}\right)^{2} + L^{ee}_{21} L^{ee}_{12} \left( 1 - L^{ee}_{22}\right)      
      \end{array}
    \right) \ .
    % \]
  \end{multline}
  The lower block analogously is
  \begin{multline}\label{eq:EI-O2-low}
    \widehat{\overline{U}_{low}\psi} \equiv \left(
      \begin{array}{cc}
        \overline{U}_{1} \psi^{1}_{2} & \overline{J}_{1}\psi^{2}_{1}\\
        \overline{J}_{1}\psi^{1}_{2} & \overline{U}_{1} \psi^{2}_{1}
      \end{array}
    \right) = \left(
      \begin{array}{cc}
        U_{1}  \psi^{1}_{2} & J_1  \psi^{2}_{1}\\[1pt]
        J_1  \psi^{1}_{2} & U_{1}  \psi^{2}_{1}
      \end{array}
    \right)  - \frac{1}{(1 + L^{ee}_{33})(1 + L^{ee}_{44}) - L^{ee}_{34} L^{ee}_{43}} \\
    \times \left(
      \begin{array}{cc}
        \left(1 + L^{ee}_{44}\right)\left( L^{ee}_{33}\right)^{2} +
        L^{ee}_{34} L^{ee}_{43} \left( 1 - L^{ee}_{33}\right) \quad &  
        L^{ee}_{34}\left( L^{ee}_{33} + L^{ee}_{44} +  L^{ee}_{33} L^{ee}_{44} - L^{ee}_{34} L^{ee}_{43}\right)  
        \\[1pt]     
        L^{ee}_{43}\left( L^{ee}_{33} + L^{ee}_{44} + L^{ee}_{33} L^{ee}_{44} - L^{ee}_{34} L^{ee}_{43}\right)    \quad & 
        \left(1 + L^{ee}_{33}\right) \left(L^{ee}_{44}\right)^{2} + L^{ee}_{43} L^{ee}_{34} \left( 1 - L^{ee}_{44}\right) 
      \end{array}
    \right) \ .
  \end{multline}
\end{widetext}
The electron-electron bubble $\psi$ and vertices $L^{ee}$ are taken
here (zero temperature) only at zero frequency.  If $\psi^{1}_{1} \neq
\psi^{2}_{2}$ we must symmetrize Eqs.~\eqref{eq:EI-O2-upp} and
\eqref{eq:EI-O2-low} by adding the Bethe-Salpeter equation with the
left averaged kernel $\mathcal{L}$ so that to keep the matrix of the
effective interaction self-adjoint.

We use the electron-hole dynamical vertex $\widehat{\overline{\Phi}}$
resulting from the simplified parquet equations and calculate the
self-energy correction to the Hartree term from Schwinger-Dyson
equation \eqref{eq:SDE-SPE}. It is in the super-matrix representation
\begin{multline}\label{eq:SDE-two_orbitals}
  \Delta \Sigma_{\iota\sigma}(i\omega_{n}) = - \frac 1\beta\sum_{m}
  \sum_{\lambda,\lambda'}\sum_{\iota',\sigma'} U^{\iota
    \iota',\sigma}_{\lambda\lambda',\sigma'}\chi^{\iota',\sigma}_{\lambda',\sigma'}
  \overline{\Phi}^{\iota'\iota,\sigma}_{\lambda'\lambda,\sigma'}(i\nu_{m})\\
  \times G_{\lambda\sigma'}(i\omega_{n} + i\nu_{m})\ .
\end{multline}
The dynamical part of the self-energy $\Delta\Sigma$ is then used to
determine the particle densities from Eq.~\eqref{eq:1P-density} so
that the static one-electron self-consistency is fulfilled.

\section{Results}\label{sec:Results}

We analyze a general two-band Hubbard model within the dynamical
mean-field theory. We set the model parameters $U_{1} = U - 2J$,
$U_{2} = U - 3J$ and $J_{1}=J_{2}=J$, which corresponds to a
spherically symmetric situation.\cite{Imada98} We choose the Hund
exchange so that all three electron interactions remain positive.  We
use the one-electron propagator from Eq.~\eqref{eq:1GF-general} with
the semi-elliptic density of states, $\rho_{1}(\epsilon) =
\rho_{2}(\epsilon) =2\pi^{-1}\sqrt{1 - \epsilon^{2}}$ the bandwidth of
which is $w = 2$.  We investigate only the spin symmetric case where
the one-electron propagators are spin-independent. The critical
interaction in the random phase approximation, separating the weak and
strong coupling regimes, then is $U_{c} =3\pi/8\approx 1.18$. It does
not depend on the number of orbitals if the Hund coupling $J=0$. We
perform all calculations at zero temperature.

\begin{figure}
  \includegraphics[width=8.5cm]
  {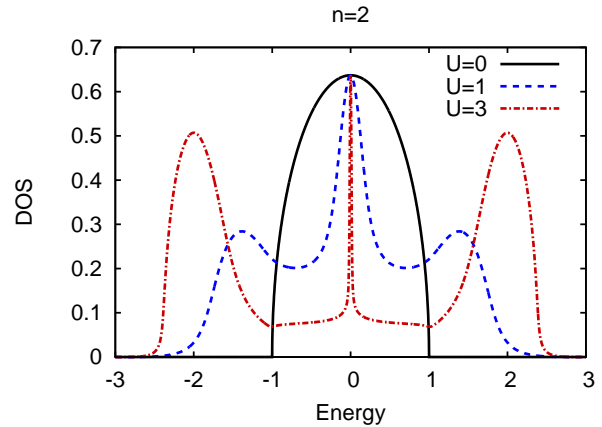}
  \caption{(Color online) Emergence of the Kondo scale in the
    two-orbital, charge, spin and and orbital symmetric model without
    Hund exchange (one electron per orbital). Semi-elliptic density of
    states.  }\label{fig:var_U}
\end{figure}

We first investigate the simplest situation of the charge-symmetric
model at half filling, that is $n=2$, with no orbital splitting
$\Delta = |E_{2} - E_{1}|/2=0$ and with vanishing Hund coupling,
$J=0$.  This situation mimics the single-orbital model. The emergence
of the Kondo scale is plotted in Fig.~\ref{fig:var_U}. The
quasiparticle peak starts to develop from the bare density of states
for interactions around the critical one from RPA, $U_{c}\approx
1.18$. The satellite Hubbard bands are well formed in the
strong-coupling regime $U>U_{c}$ and lie slightly beyond the atomic
values $\pm U/2$.  The Kondo scale remains exponentially small but
non-zero even for large interaction strengths. The Kondo asymptotics
for the semi-elliptic density of states is $a\propto \exp\{-
4U\rho_{0}/3\}$, where $\rho_{0}$ is the density of states at the
Fermi energy.  This result is in discrepancy with general confidence
that the Hubbard model at half-filling undergoes in the paramagnetic
phase a Mott-Hubbard metal insulator transition at a finite
interaction strength.  The existence of a metal-insulator transition
in DMFT is concluded from extrapolations of low-temperature
Monte-Carlo\cite{Georges96} or NRG\cite{Bulla01,Bulla08}
data. Although there is no doubt about the existence of an insulating
solution for very large interaction strengths matching the atomic
limit, numerical methods cannot presently exclude the existence of a
metallic solution with a very narrow (below numerical resolution)
quasiparticle peak. The existence of a sharp metal-insulator
transition has not yet been rigorously proved and there is at present
neither a consistent solution at the critical point of a transition
nor there is an analytic formula for the vanishing width of the
quasiparticle peak. If there is a metal-insulator transition, the
metallic side of the transition cannot be a Fermi liquid. The
low-frequency asymptotics of the electron-hole correlation function in
Fermi liquid is due to sum rules of order $\omega^{-1}$.  Such a
behavior is then incompatible with Bethe-Salpeter equations that have
to be fulfilled self-consistently in the electron-hole and
electron-electron scattering channels by an exact solution. Only
integrable singularities can emerge in an exact solution and Fermi
liquid hence does not allow for a divergence in the electron-hole
correlation function expected at the Mott-Hubbard metal-insulator
transition. In this respect the simplified parquet approximation is
consistent with the exact two-particle behavior.

Nonexistence of a sharp metal-insulator transition in the simplified
parquet equations is not caused by approximations made in this
approach but rather it is a consequence of a Fermi-liquid solution
fulfilling self-consistently Bethe-Salpeter equations simultaneously
in the electron-hole and electron-electron scattering channels. There
is an insulating paramagnetic state in parquet-based theories for $U >
U_{c}$ (critical interaction of the RPA pole) being a superposition of
Hartree polarized solutions, but it coexists with the metallic one in
the strong-coupling regime and has a higher total energy. A
metal-insulator transition in the Fermi liquid can either be connected
with a breaking of a global symmetry and/or can be noncritical as we
discuss later for orbital splitting in a crystal field.
%
% \begin{figure}
%   \includegraphics[width=8.5cm]
%   {var_U_lorentz.eps}
%   \caption{(Color online) Emergence of the Kondo scale in the same
%   setting as in Fig.~\ref{fig:var_U} for the lorentzian density of
%   states (single impurity Anderson
%   model). }\label{fig:var_U_Lorentz}
% \end{figure}
%

% We can compare emergence of the Kondo scale in the strong-coupling
% regime of the Hubbard model obtained within the dynamical mean-field
% approximation with a semi-elliptic density of states and the Kondo
% asymptotics in the single-impurity Anderson model, where we use the
% one-electron propagator from Eq.~\eqref{eq:1GF-SIAM} with
% $\Delta(i\omega_{n}) = i\Delta \text{sign} (\omega_{n})$,
% Fig.~\ref{fig:var_U_Lorentz}. The critical RPA interaction strength
% is $U_{c}=\pi$ and the Kondo regime sets in for larger interaction
% strengths. The Kondo exponential scale in the limit of very strong
% interaction $U\to\infty$ is $a\propto \exp\{- U\rho_{0}\}$. The
% satellite bands are broader and less peaked and centered closer to
% the atomic values $\pm U/2$ than in the case of the semi-elliptic
% band. It is clear that the central quasi-particle peak in DMFT is
% for a given interaction strength narrower than in SIAM, as indicates
% the asymptotic solution for the Kondo scale and is also seen in NRG
% calculations in the single-orbital models.\cite{Bulla01}

\begin{figure}
  \includegraphics[width=8.5cm]
  {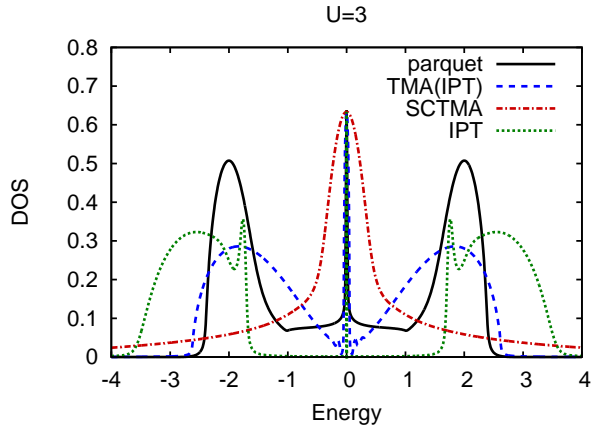}
  \caption{(Color online) Spectral function for the symmetric
    situation calculated in various diagrammatic approximations:
    simplified parquets, self-consistent T-matrix (SCTMA), iterated
    perturbation theory (IPT), and T-matrix with IPT
    propagators. }\label{fig:U3}
\end{figure}

\begin{figure}
  \includegraphics[width=8.5cm]
  {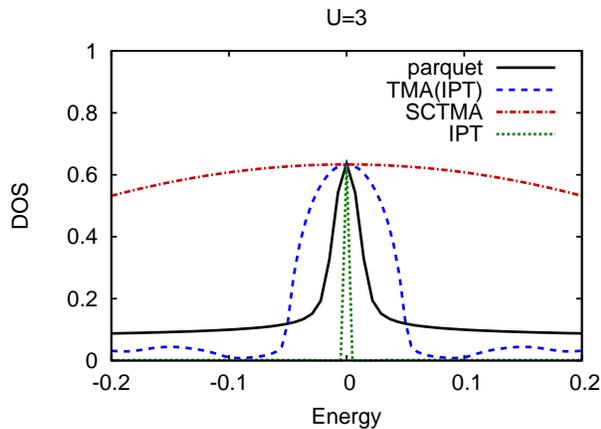}
  \caption{(Color online) Detailed structure of the quasiparticle peak
    for the approximations from
    Fig.~\ref{fig:U3}.}\label{fig:U3_detail}
\end{figure}

We compared the spectral function calculated from the simplified
parquet equations with other diagrammatic approximations based on
summations of Feynman diagrams for multiple two-particle
scatterings. We chose a value of the interaction strength $U=3$ so
that to be in the strong-coupling regime where renormalizations of the
perturbation expansion are necessary. We used the spectral function
for the self-consistent T-matrix, iterated perturbation theory in
second order (IPT) and for T-matrix with IPT (topological)
self-consistency, that is, the local renormalized propagator of the
Hubbard model $\mathcal{G}$ is defined as $\mathcal{G}^{-1}(\omega) =
G^{-1}(\omega - \Sigma(\omega)) + \Sigma(\omega)$.  The results of the
comparison are plotted in Fig.~\ref{fig:U3}.  We can see that except
for the fully self-consistent T-matrix, the chosen three
approximations display the expected three-peak structure. They differ,
however, in the way the satellite bands are attached to the central
quasiparticle peak. The simplified parquet approximation keeps a small
finite density of states between the central and satellite bands. It
means that this approximation allows for an energy transfer between
the central and the satellite bands. Approximations based on the IPT
self-consistency develop a quasi-gap between the low and high energy
states and no energy exchange occurs. The satellite bands are less
spread in the parquet approximation than in the IPT-based ones. The
detailed shape of the central quasi-particle peak in these
approximations is magnified in Fig.~\ref{fig:U3_detail}.  The most
narrow is the central peak for IPT suggesting vicinity of a
metal-insulator transition with zero width of the quasiparticle
peak. The parquet-based approximations, on the other hand, slow down
the tendency toward a metal-insulator transition. They actually screen
it, since there may be no critical metal-insulator transition in these
theories as discussed above.


\begin{figure}
  \includegraphics[width=8.5cm]
  {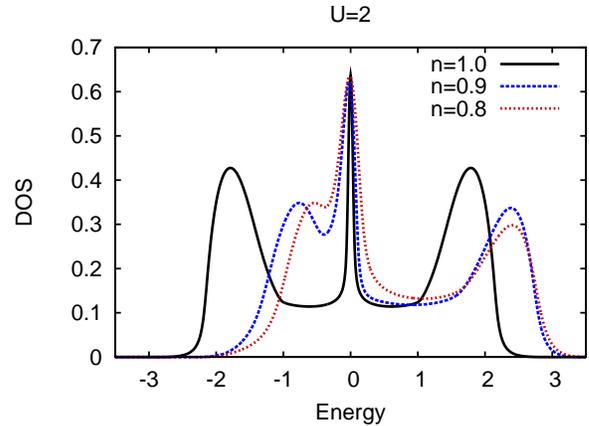}
  \caption{(Color online) Spectral function in charge asymmetric
    situations for various filings $n$, averaged number of electrons
    per orbital.}\label{fig:var_fill}
\end{figure}

The simplified parquet equations show best agreement with the exact
solution of the single-impurity Anderson model at half filling, where
the density of states at the Fermi surface is fixed. The same holds
also for other fillings at intermediate and strong coupling calculated
within the dynamical mean-field approximation. Apart from the
charge-symmetric situation the simplified parquet equations reproduce
qualitatively well the low and high frequency features of the spectral
function, although the width of the quasiparticle peak and the Kondo
temperature derived from it do not follow as closely the exact
solution for the flat-band model as for the half-filled
case.\cite{Janis08} The spectral function of a strong-coupling
solution changes when receding from half filling so that the central
quasiparticle peak broadens and slightly moves below the Fermi
surface.  The lower satellite band moves toward the Fermi energy. The
lower band and the central peak eventually merge, the quasiparticle
peak is absorbed by the satellite one, when the electron-hole
asymmetry becomes prominent. This feature seems to be universal for
all theories and is demonstrated in our approach in
Fig.~\ref{fig:var_fill}.

\begin{figure}
  \includegraphics[width=8.5cm]
  {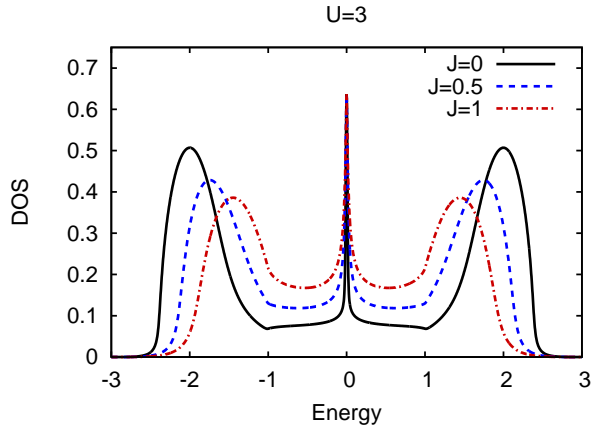}
  \caption{(Color online) Effect of the Hund exchange coupling on the
    spectral function.}\label{fig:var_J}
\end{figure}

We hitherto have shown the results for the two-orbital model without
Hund exchange and orbital splitting, that is for $J=0$ and $\Delta =
0$. Such a situation is close to the single-orbital model as we
demonstrated.  The most important change in the model with nonzero
Hund coupling is a splitting of the interaction strength, that is,
$U_{1}= U - 2J$ and $U_{2}= U - 3J$. It means that the interaction
strength controlling the approach to the critical point in the
electron-hole channel is the Hubbard $U$ between different spins on
the same orbital. The Hund exchange decreases the critical interaction
at which the electron-hole vertex diverges in the bare perturbation
theory. It would then narrow the width of the Kondo resonant peak at a
fixed renormalized Coulomb repulsion $\overline{U}$. In the real
situation we, however, do not fix the renormalized but rather the bare
Coulomb repulsion.  We find that though larger Hund exchanges lead to
higher values of the averaged kernels $L^{ee}$, as we can see from
Eqs.~\eqref{eq:EI-O2-upp} and \eqref{eq:EI-O2-low}, they
simultaneously cause decrease in the renormalized repulsions
$\overline{U}_{\alpha}$ at a fixed bare Coulomb interaction.
Consequently, the weight of the central quasiparticle peak in the
self-energy decreases. We plotted the impact of the Hund coupling on
the spectral function in Fig.~\ref{fig:var_J}.  We observe that the
width of the Kondo peak shrinks but only on a narrower frequency
interval near the Fermi energy.  The Hund coupling at a fixed bare
Coulomb repulsion then leads to filling of the states between the
central peak and the satellite bands. Further, the satellite bands are
positioned closer to the central one and become less pronounced with
the increasing Hund exchange.

\begin{figure}
  \includegraphics[width=13cm]
  {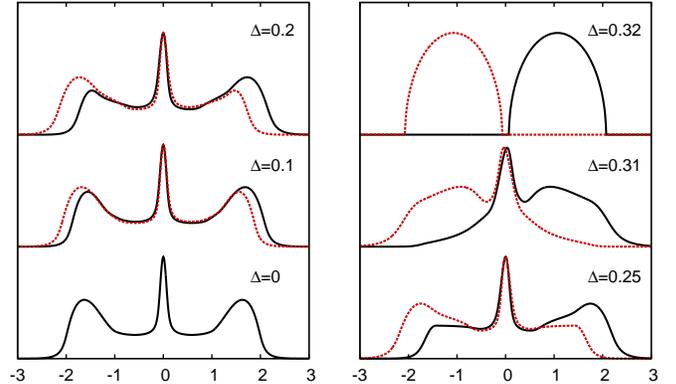}
  \caption{(Color online) Orbital splitting and a metal-insulator
    transition in the half-filled case for the semi-elliptic density
    of states and $U=1.5$. }\label{fig:MIT-orbital}
\end{figure}

\begin{figure}
  \includegraphics[width=13cm]
  {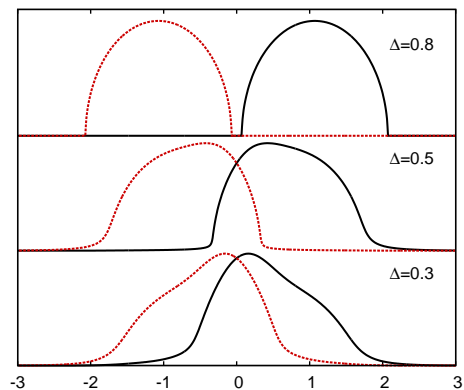}
  \caption{(Color online) Orbital splitting and a metal-insulator
    transition in weak coupling, $U=0.5$.}\label{fig:MIT-orbital_weak}
\end{figure}

The most interesting on the two-orbital model are changes in the
spectral function due to a broken orbital symmetry induced by a
crystal field. That is, when the one-electron propagators are
different on different orbitals. We achieve this here by splitting the
centers of the two bands. We set $E_{1}= - E_{2}= \Delta$, but keep
the same band structure, $\rho_{1}(\epsilon) =
\rho_{2}(\epsilon)$. The effect of the orbital splitting depends on
whether we have a finite or infinite bandwidth of macroscopically
occupied energies.  Orbital splitting leads in the former case to a
metal-insulator transition unlike the latter one. The genesis of a
metal-insulator transition in the model with the semi-elliptic density
of states is plotted in Fig.~\ref{fig:MIT-orbital}. We chose a
moderate interaction strength, $U = 1.5$, so that to be able to follow
the process more closely. When we increase the orbital splitting
parameter $\Delta$ there is almost no observable change at the Fermi
energy with a pronounced quasiparticle peak. It slightly splits into
two but stays at the Fermi energy.  More apparent is the change caused
by orbital splitting in the bulk of the energy band where the states
are rearranged so that the orbital polarization is enhanced. Such a
response on the band splitting effectively pushes the system out from
the critical region of the RPA pole.  With the increasing band
splitting the dynamical renormalization of the band structure, apart
from the vicinity of the Fermi energy, becomes less important and the
solution approaches the Hartree one. The process continues up to the
point when one of the orbital saturates and the other is emptied. The
insulating solution is then that of the Hartree approximation with no
dynamical correlations. Changes in the quasiparticle peak, split and a
broadening, are observed only close to the transition. They vanish
practically at the transition point. There are, however, no singular
two-particle functions at the transition point, the metal-insulator
transition is hence noncritical. It is worth noting that the Kondo
quasiparticle peaks exist in a pronounced way only in the
strong-coupling regime, cf. Fig.~\ref{fig:MIT-orbital_weak}.

\begin{figure}
  \includegraphics[width=8.7cm]
  {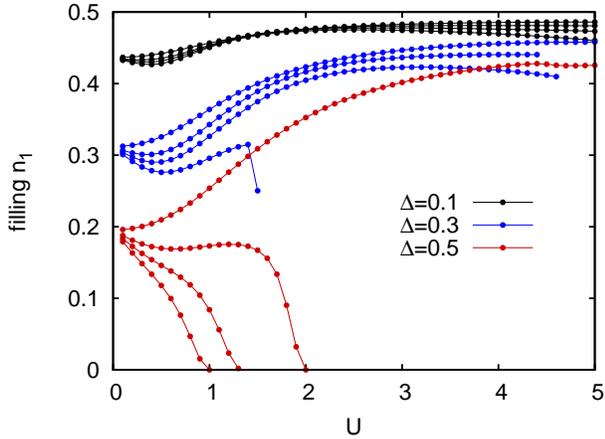}
  \caption{(Color online) Filling per spin of the upper orbital as a
    function of the bare interaction strength $U$. Different curves
    for a given split $2\Delta$ correspond to different $J$ values
    from bottom to top $J/U=0, 0.05, 0.1, 0.2$. A metal-insulator
    transition ($n_{1}=0$) takes place for values of $J/U < 0.2$.  The
    Kondo regime, vicinity of the Mott-Hubbard transition
    ($n_{1}\approx 1/2$) is reached only for larger values of
    $J/U$. }\label{fig:fill_U_J}
\end{figure}

We further studied the effect of the Hund exchange on the orbital
metal-insulator transition. We set the parameters of the model so that
we could compare our results with the existing Monte-Carlo simulations
of Ref.~\onlinecite{Werner07}.  We chose three values of the bare
orbital splitting $2\Delta = 0.2, 0.4, 1.0$. For each value of the
band splitting we fixed four ratios of the Hund coupling $J/U = 0,
0.05, 0.1, 0.2$ and plotted the filling (per spin) of the upper
orbital $n_{1}$ as a function of the bare interaction at half filling
$n_{1} + n_{2}= 1$ in Fig.~\ref{fig:fill_U_J}.  We observe that small
Hund couplings do not prevent the system from falling into the
orbitally polarized state as demonstrated in
Fig.~\ref{fig:MIT-orbital}. For large $\Delta$ the occupation of the
upper orbital decreases practically monotonically with the increasing
interaction. For smaller splittings the filling $n_{1}$ increases and
may get to the Kondo regime at intermediate interaction strengths
where both orbitals are almost equally occupied ($n_{1}\approx n_{2}
\approx 0.5$).  Unlike the Monte-Carlo data, we do not reach, however,
the Mott insulating phase from the Fermi-liquid solution. If $J/U <
0.2$ the system does not stay in the Kondo regime for ever. Strong
electron repulsion increases orbital polarization due to the Hartree
shift and the system eventually goes over into a polarized insulator
$n_{1}=0$. For ratios $J/U>0.2$ the Hartree term starts to act against
the crystal field and diminishes orbital polarization. The filling
$n_{1}$ then asymptotically approaches half filling.  Apart from
nonexistence of the Mott transition these effects of the Hund coupling
on the occupation of orbitals in a crystal field are described by the
simplified parquet equations in a good agreement with the Monte-Carlo
simulations. It is worth mentioning that the Kondo resonance for $J/U
< 0.2$ survives to a certain value of the interaction strength beyond
which it is rather abruptly suppressed, similarly to the scenario from
Fig.~\ref{fig:MIT-orbital}. Vanishing of the Kondo resonance is
followed by a sharp saturation of orbital polarization. The numerical
solution then becomes unstable, since the metallic phase in the
extreme Kondo regime and the orbitally polarized insulating solution
coexist with almost the same energies.

\begin{figure}
  \includegraphics[width=13cm]
  {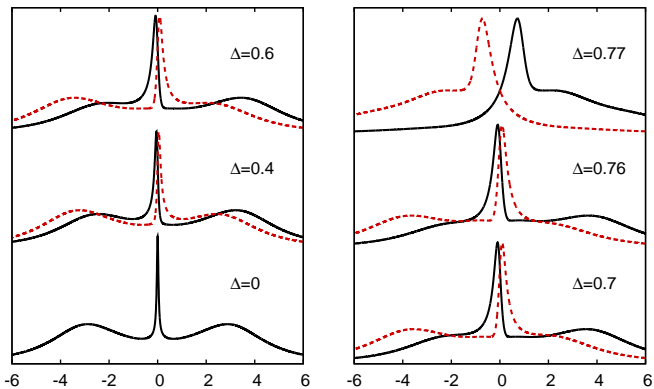}
  \caption{(Color online) Orbital splitting in the single-impurity
    Anderson model for $U=5$. No metal-insulator transition takes
    place. }\label{fig:MIT-orbital_Lorentz}
\end{figure}

The scenario of orbital splitting changes if the bare energy bands are
infinite. When an orbital symmetry breaking is switched on, the system
with infinite band-widths behaves for small splittings analogously to
the model with finite energy bands. The changes take place in the
bulk, away from the Fermi energy so that the orbital splitting is
enhanced, cf. Fig.~\ref{fig:MIT-orbital_Lorentz}. Similarly there are
almost no observable changes on the quasiparticle peak. The system is
slowly pushed away from the critical region of the RPA transition by
increasing the orbital splitting and approaches orbital
saturation. But due to the infinite band width, there is no transition
to the Hartree insulator and the density of states at the Fermi energy
remains nonzero for all interaction strengths. The Kondo peaks do not
survive to infinite interaction.  A sudden macroscopic cleft breaks in
between the central quasiparticle peaks of different orbitals at a
finite interaction strength. The split Kondo peaks are then soon
absorbed by the bulk energy bands. The solution approaches a
weak-coupling (Hartree-Fock) state with almost separated energy bands
centered around energies $\pm (\Delta + Um_{o}/2)$, where $m_{o}$ is
the orbital moment.

\section{Discussion and conclusions}\label{sec:Conclusions}

Dynamical effects of strong electron correlations in impurity models
manifest themselves in a three-peak structure of the spectral function
with a narrow central quasiparticle peak near the Fermi energy and two
satellite bands centered around eigenvalues of the local part of the
interacting Hamiltonian. There is no exact solution proving such a
picture, but a number of approximate numerical and analytic-numerical
solutions confirm this behavior of the spectral function. There are a
number of such approximate schemes for the simplest single-orbital
impurity and DMFT models. Extensions of approximations showing the
expected three-peak spectral function in strong coupling to
multi-orbital models face a considerable increase of complexity in the
strong-coupling regime. Monte-Carlo simulations as well as the
numerical renormalization group applied on multi-orbital models need
to resort to additional approximations to reach sufficient accuracy of
dynamical and spectral properties.

Most of the existing approximations on dynamical properties of
strongly-correlated electron systems make directly accessible only
one-electron properties. Only few of them, even in single-orbital
models, are able to approach two-particle quantities and in
particular, to control two-particle divergencies that develop in
Bethe-Salpeter equations.  We extended in this paper an approximation
based on parquet equations controlling the RPA singularity in the
electron-hole scattering channel to a general multi-orbital model of
correlated electrons. Approximations used to simplify complexity of
the full parquet approach resulting from a separation of large and
small dynamical fluctuations in the critical region of the RPA pole
allow us to reproduce the expected three peak structure of the
spectral function of SIAM in the Kondo regime. An extension of this
approximation from single to multi-orbital models proved to be
manageable with only a moderate increase in numerical requirements. An
$l$-orbital model is in the most general case described by
$4l^{2}\times 4l^{2}$ matrices and the numerical expenses increase
polynomially with the number of orbitals. In real situations
symmetries and conservation laws further reduce complexity of the
simplified parquet equations.

The impurity solver for the dynamical mean-field theory we built up
within the parquet approximation suits best for describing the
dynamical and spectral properties of the metallic (Fermi liquid) phase
from weak up to moderately strong electron correlations. No
self-consistency is needed in the weak-coupling regime below the
critical interaction strength $U_{c}$ at which the RPA vertex
diverges. The parquet self-consistency is indispensable in
intermediate ($U\approx U_{c}$) and strong ($U > U_{c}$) coupling. The
numerical procedure for the solution of the simplified parquet
equations gets unstable for large interaction strengths. We were able
to reach a stable metallic solution up to $U \approx 5 U_{c}$. Unlike
other approaches such NRG, QMC or IPT this approximation does not
allow for a \textit{critical} metal-insulator transition without a
symmetry breaking in Fermi liquid. This is a general feature of all
impurity solutions where Bethe-Salpeter equations in the electron-hole
and electron-electron channels should be obeyed simultaneously in a
self-consistent manner. Parquet equations are hence unable to describe
a Mott insulator without a symmetry breaking. The insulating solution
obtained within the local parquet approach is a superposition of
fluctuations-free Hartree saturated solutions and the transition to
such a state is noncritical.

The simplified parquet equations presented in this paper are not
quantitatively as accurate as numerically exact solutions, but they
offer a numerically manageable scheme with an analytic control of
two-particle singularities in Bethe-Salpeter equations.  The
self-energy and the spectral function calculated within this
approximation obey Fermi-liquid rules in the metallic phase and
produce the typical three-peak structure in the density of states in
the strong-coupling regime. The width of the central quasiparticle
peak is exponentially small for large interaction strengths. This
approximation may not reproduce the exact asymptotics of the Kondo
scale and temperature in the single-impurity model as well as
e.~g. the local moment approach, but unlike the latter it does not
show any unphysical behavior and does not use auxiliary fitting
parameters. The approximation based on the simplified parquet
equations is formulated in such a way that it can directly be applied
not only to impurity models but to lattice models within the dynamical
mean-field approach as an impurity solver.

The impurity solver based on the simplified parquet equations is a
semi-analytic theory for local spectral properties formulated in real
frequencies with a full control of analytic properties of the
approximation. From this reason the approximation best suits to the
calculation of spectral properties and the dynamical self-energy
directly accessible from a perturbation theory. Its unique feature is
that it offers analytic control of two-particle singularities in the
Bethe-Salpeter equations. The simplified parquet approximation is well
suited to problems involving two-particle dynamical, and in particular
critical, functions. The approximation is less suited to global
thermodynamic quantities being averages of dynamical functions over
frequencies. An extension to finite temperatures does not represent a
problem and the effective interactions can be calculated also at
nonzero temperatures as discussed in this paper. The quantitative
reliability of the simplified parquet equations decreases with the
reduced influence of the critical behavior of the vertex function in
the vicinity of the RPA pole (Kondo regime). Only in the critical
region of a pole in the two-particle vertex we can rely on a
separation of large and small dynamical fluctuations, that is, a
decoupling of singular and non-singular functions.

To our knowledge, the simplified parquet equations is the only scheme
that uses a two-particle self-consistency to control the Kondo
asymptotics and criticality of the RPA pole. There are attempts to use
parquet equations for renormalizations of the perturbation expansion,
but they are based on the full one- and two-particle
self-consistency.\cite{Bickers04,Yang09} It is, however, important to
realize that the full dynamical one-particle self-consistency impedes
control of two-particle divergencies and the Kondo scale in the
spectral function is no longer correctly reproduced. Moreover, the
full dynamical one-electron self-consistency ultimately leads to
smearing of the satellite bands.\cite{Janis07a} It is hence of
importance that the parquet equations are solved with the Hartree
propagators so that to reproduce the three-peak spectral function with
the correctly scaled central quasiparticle peak.

The extension of the simplified parquet equations was used in this
paper as an impurity solver for the two-band Hubbard model with Hund
coupling and orbital splitting. We showed emergence of the Kondo
quasiparticle peak with the increasing interaction strength at half
filling. The quasiparticle peak survives only close to the
electron-hole symmetry where multiple electron-hole scatterings are
dominant. When deviations from this symmetry are significant the
quasiparticle peak is absorbed by one of the satellite bands. We also
demonstrated how a metal-insulator transition induced by a crystal
field takes place in a model with finite energy bands. The parquet
equations do not allow for a critical transition in the local Fermi
liquid. We showed that the orbitally induced metal-insulator
transition is noncritical, but with a Kondo behavior almost up to the
transition point, if the interaction is sufficiently strong.

Last but not least, the approximate impurity solver we built up on
parquet equations uses the formalism of Green functions and can
conveniently be combined with ab-initio computational schemes to
assess correlation effects in materials with a nontrivial atomic
structure. It should be suitable in particular for metallic systems
where we expect Kondo-like behavior or strong valence fluctuations.

\acknowledgments \noindent We thank V. Drchal and A. B. Shick for
valuable discussions and helpful comments. This research was carried
out within project AV0Z10100520 of the Academy of Sciences of the
Czech Republic and supported in part by Grant No.  202/07/J047 of the
Czech Science Foundation.


\begin{thebibliography}{99}
\bibitem{Anderson61} P. W. Anderson, \newblock Phys. Rev. {\bf 124},
  41 (196).

\bibitem{Kondo64} J. Kondo,\newblock Prog. Theor. Phys. {\bf 32}, 37
  (1964).

\bibitem{Tsvelik83} A. M. Tsvelik and P. B. Wiegmann, \newblock Adv.
  Phys. {\bf 32}, 453 (1983).
  
\bibitem{Borda03} L. Borda, G. Zarand, W. Hofstetter, B. I. Halperin,
  and J. von Delft, \newblock Phys. Rev. Lett. \textbf{90}, 026602
  (2003).

\bibitem{Choi05} M. S. Choi, R. L\'opez, and R. Aguado, \newblock
  Phys. Rev. Lett. \textbf{95}, 067204 (2005).

\bibitem{Georges96} A. Georges, G. Kotliar, W. Krauth, and M.
  Rozenberg, \newblock Rev. Mod. Phys. {\bf 68}, 13 (1996).

\bibitem{Fye88} R. M. Fye and J. E. Hirsch, \newblock Phys. Rev.  B
  {\bf 38}, 433 (1988).

\bibitem{Rubtsov05} A. N. Rubtsov, V. V. Savkin, and
  A. I. Lichtenstein, \newblock Phys. Rev. B \textbf{72}, 035122
  /2005).

\bibitem{Werner06a} P. Werner, A. Comanac, L. de' Medici, M. Troyer,
  and A. J. Millis, \newblock Phys. Rev. Lett. \textbf{97}, 076405
  (2006).

\bibitem{Caffarel94} M. Caffarel and W. Krauth, \newblock
  Phys. Rev. Lett. \textbf{72}, 1545 (1994).

\bibitem{Wilson75} K. G. Wilson, \newblock Rev. Mod. Phys.  {\bf 47},
  773 (1975).

\bibitem{Krishnamurthy80} H. R. Krishna-murthy, J. W. Wilkins, and K.
  G. Wilson, \newblock Phys. Rev. B {\bf 21}, 1003, 1044 (1980).
 
\bibitem{Costi94} T. A. Costi, A. C. Hewson, and V. Zlati\'c,
  \newblock J. Phys.: Condens. Matter {\bf 6}, 2519 (1994).

\bibitem{Bulla98} R. Bulla, A. C. Hewson, and T. Pruschke, \newblock
  J. Phys.: Condens. Matter {\bf 10}, 8365 (1998).
 
\bibitem{Hewson93} A. C. Hewson, \newblock \textit{The Kondo Problem
    to Heavy Fermions} (Cambridge University Press, Cambridge 1993).

\bibitem{Hewson93a} A. C. Hewson, \newblock Phys. Rev. Lett.
  \textbf{70}, 4007 (1993).
 
\bibitem{Keiter70} H. Keiter and J. C.  Kimball, \newblock Phys.  Rev.
  Lett. {\bf 25}, 672 (1970).

\bibitem{Pruschke89} T. Pruschke and N. Grewe, \newblock Z.  Physik B
  {\bf 74}, 439 (1989).

\bibitem{Bickers89} N. E. Bickers and D. J. Scalapino, \newblock Ann.
  Phys. (N. Y.) {\bf 193}, 206 (19989).

\bibitem{Logan98} D. E. Logan, M. P. Eastwood, and M. A. Tusch,
  \newblock J. Phys.: Condens. Matter \textbf{10}, 2673 (1998).
 
\bibitem{Janis07a} V. Jani\v s and P. Augustinsk\'y, \newblock Phys.
  Rev. B {\bf 75}, 165108 (2007).

\bibitem{Janis08} V. Jani\v s and P. Augustinsk\'y, \newblock Phys.
  Rev. B {\bf 77}, 085106 (2007).
  
\bibitem{Ogievetski83} E. Ogievetski, A,. M. Tsvelik, and
  P. B. Wiegmann, \newblock J. Phys. C: Solid State Phys. \textbf{16},
  L797 (1983).

\bibitem{Bonca93} J. Bon\v{c}a and J. E. Gubernatis, \newblock
  Phys. Rev. B \textbf{47}, 13137 (1993).

\bibitem{Kotliar06} G. Kotliar, S. Y. Savrasov, K. Haule,
  V. S. Oudovenko, and O. Parcollet, \newblock
  Rev. Mod. Phys. \textbf{78} 865 (2006).

\bibitem{Werner06b} P. Werner and A. J. Millis, \newblock Phys. Rev. B
  \textbf{74}, 155107 (2006).

\bibitem{Liebsch05} A. Liebsch, \newblock
  Phys. Rev. Lett. \textbf{95}, 116402 (2005).

\bibitem{Inaba05} K. Inaba, A. Koga, S, Suga, and N. Kawakami,
  \newblock J. Phys. Soc. Jpn. \textbf{74}, 2393 (2005).

\bibitem{Galpin05} M. R. Galpin, D. E. Logan, and H. R. Krishnamurthy,
  \newblock Phys. Rev. Lett. \textbf{94}, 186406 (2005).

\bibitem{Bulla08} R. Bulla, T. Costi, and T. Pruschke, \newblock
  Rev. Mod. Phys. \textbf{80}, 395 (2008).

\bibitem{Nishikawa10} Y. Nishikawa, D. J. G. Crow, and A. C. Hewson,
  \newblock e-print arXiv:1005.5113.

\bibitem{Drchal05} V. Drchal, V. Jani\v{s}, J. Kudrnovsk\'y,
  V. S. Oudovenko, X. Dai, K. Haule, and G. Kotliar, \newblock
  J. Phys.: Condens. Matter \textbf{17}, 61 (2005).

\bibitem{Galpin09} M. R. Galpin, A. B. Gilbert, and D. E. Logan,
  \newblock J. Phys.: Condens. Matter \textbf{21}, 375602 (2009).

\bibitem{Kauch09} A. Kauch and K. Byczuk, \newblock e-print
  arXiv:0912.4278.
 
\bibitem{Janis99b} V. Jani\v s, Phys. Rev. B \textbf{60}, 11345
  (1999).

\bibitem{Baym62} G. Baym, \newblock Phys. Rev. {\bf 127}, 1391 (1962).

\bibitem{Janis09} V. Jani\v{s} and M. Ringel, \newblock Acta
  Phys. Polon. A \textbf{115}, 30 (2009).

\bibitem{Imada98} M. Imada, A. Fujimori, and Y. Tokura, \newblock
  Rev. Mod. Phys. \textbf{70}, 1039 (1998).
 
\bibitem{Bulla01} R. Bulla, T. A. Costi, and D. Vollhardt, \newblock
  Phys. Rev.  B \textbf{64}, 045103 (2001).

\bibitem{Werner07} P. Werner and A. J. Millis, \newblock
  Phys. Rev. Lett.  \textbf{99}, 126405 (2007).

\bibitem{Bickers04} N. E. Bickers, in \textit{Theoretical Methods for
    Strongly Correlated Electrons}, edited by D. Senechal,
  A. Tremblay, and C. Bourbonnais (Springer-Verlag, New York, 2004),
  p. 237.

\bibitem{Yang09} S. X. Yang at al, \newblock Phys. Rev. E \textbf{80},
  046706 (2009).
 
\end{thebibliography}
\end{document}